\pdfoutput=1 %arXiv admin added the following line
\documentclass[aps,pre,floatfix]{revtex4}
\usepackage{graphicx}
\usepackage{color}
\usepackage{floatflt}
\usepackage{lscape}
\usepackage{wrapfig}
\usepackage{setspace}

\begin{document}
\normalsize
\title{From differential equations to Boolean networks: A Case Study in modeling regulatory networks}
\author{Maria Davidich\footnote{Corresponding author: davidich@itp.uni-bremen.de, Tel: 49-421-218-3195, Fax. +49-421-218-9104}}
\author{Stefan Bornholdt}
\affiliation {Institute for Theoretical Physics, University of
Bremen}
\address {Otto-Hahn-Allee, NW1, D-28359 Bremen
Germany}
\email{ davidich@itp.uni-bremen.de,  bornholdt@itp.uni-bremen.de}
\bibliographystyle{apalike}
\begin{abstract}
Methods of modeling cellular regulatory networks as diverse as
differential equations and Boolean networks co-exist, however, without
any closer correspondence to each other. With the example system
of the fission yeast cell cycle control network, we here set the two
approaches in relation to each other. We find that the Boolean network
can be formulated as a specific coarse-grained limit of the more
detailed differential network model for this system.
This lays the mathematical foundation on which Boolean networks
can be applied to biological regulatory networks in a controlled way.
\end{abstract}
\maketitle
\section{Introduction}
The task of molecular cell biology is to comprehend the control of
cellular processes of living cells encoded in the genome of the
cell. These cellular processes are guided by sophisticated networks
of interactions between the macromolecules of the cell as
proteins, nucleic acids and polysaccharides. Their complexes and
structures define the unique properties that enable them to perform
the functions of the cell as, for example, catalysis of chemical
transformations, production of movement, and heredity.
The complexity of these processes demands not only advanced
experimental techniques, but also adequate mathematical and
computational models for understanding them (Gunsalus et al., 2005; Riel, 2006).

Today, there are different methods for modeling the complex networks
of biochemical interactions, ranging from master equations based on
first principles (Monte-Carlo
method) (Gillespie, 1976; Gillespie, 1977), ordinary differential
equations (ODE) (Aguda, 2006; Chen et al., 2000; Novak and Tyson, 1993; Novak and Tyson, 1997; Novak et al., 2001;
Novak and Tyson, 2004; Sveiczer et al., 2000; Tyson et al., 2001; Tyson et al., 2003),
stochastic differential equations (Fokker-Plank equations), all the
way to Boolean networks (Albert and Othmer, 2003; Bornholdt, 2005; Sanchez and Thieffry, 2001; Thomas et al., 1995).

Among these methods, a most popular approach to modeling biochemical
networks is via differential equations, based on the known chemical
kinetics and successfully applied to describing numerous processes in
living organisms (Chen et al., 2000; Novak and Tyson, 1993; Novak and Tyson, 1997; Novak et al., 2001;
Novak and Tyson, 2004; Tyson et al., 2002). To build an ODE model, one starts with a
schematic diagram representing the known interactions between
components. Then this diagram is converted into a set of
differential and algebraic equations. The full ODE model then
consists of this set of rate equations, plus a set of parameter
values as well as a set of initial conditions. The solutions of the
ODEs give the time-dependence of each component of the system. In
practice these solutions depend on rather detailed knowledge about
all reactions and kinetic parameters.

In studies where prediction of exact reaction times is not an issue,
simpler models than ODE and less parameters may be necessary for
predicting the course of events in a regulatory network. For example,
relevant features of cell commitment, cell cycle progression, and
cell differentiation are already described in terms of a sequence of
regulatory events (Albert and Othmer, 2003; Braunewell and Bornholdt, 2006;
Davidich and Bornholdt, 2008; Li et al., 2004; Sanchez et al., 2001). In such cases, the much simpler modeling framework of Boolean networks may be a suitable method (Kaufmann, 1969; Thomas, 1973).
Constructing a Boolean network model starts from a wiring diagram of
interactions between biochemical elements as well, but no kinetic
details are needed. Interactions are classified into just two
classes, activation or inhibition, as well as the concentration
levels being reduced to just an ON or OFF state.

Despite their extreme simplicity, such Boolean models are able to
reproduce regulatory sequences, for example, models of genetic networks of {\em A.\ thaliana},
(Espinosa-Soto et al., 2004; Mendoza et al., 1999; Thum et al., 2003), the cell-cycle
network of {\em S. cerevisiae} (Li et al., 2004), the mammalian
cell-cycle (Faure et al., 2006), and the segment polarity gene network
in {\em D.\ melanogaster} (Albert and Othmer, 2003; Sanchez et al., 2001).
These examples show that the Boolean network approach provides
reliable results for different organisms.

The two diverse methods just summarized (ODE and Boolean networks)
are both based on the same ``wiring'' diagram of interactions
between the components, however, use much different amounts
of information about these interactions.
This poses the interesting question how these two methods
are related to each other.
A first correspondence between Boolean network and ordinary
differential equations has been drawn by Glass and collaborators
(Glass and Kauffman, 1973; Glass and Hill, 1998) who explored the
relationship between a class of non-linear equations representing
biochemical control networks and homologous switching networks.
They argued that such a correspondence can be achieved with the
following requirements: 1) Rates of reactions are described by monotonic
sigmoidal functions having distinct upper and lower asymptotes. 2)
The parameters must be defined to match the upper or lower asymptote.
3) The target control function must correspond to the maximal or
basal rate of biochemical processes. They subsequently demonstrated
that there is a large variety of such functions as, e.g., the Heaviside
function, the error function, or the Hill function defined for
positive arguments.

This leads to a mapping between asymptotical solutions of the
ODE system and the Boolean system, while omitting the exact
way of transitions between dynamical states.

In this paper, we further explore the correspondence between
ODE and Boolean network models considering a specific
biological system and demonstrate how a working
Boolean model can be derived in terms of a mathematically
well defined coarse-grained limit of an underlying ODE model.
As our working example we choose the fission yeast cell-cycle
control network ({\em Schizosaccharomyces Pombe}).
The division cell-cycle consists of four phases G1--S--G2--M, during
which DNA is replicated and the cell divides itself into two cells.
The main role is played by "Cyclin-Dependent-Kinases" (CDKs)
and cyclins that bind to CDKs to form complexes. CDKs, while
being present at all times, can only be active in complexes with
cyclins. Cyclins are synthesized or degraded depending
on other regulatory activities. Another important participant of the
process is an enzyme complex called the "Anaphase-Promoting Complex"
(APC), which targets cyclins for degradation. In summary,  "to
understand the molecular control of cell reproduction is to
understand the regulation of CDK and APC activities"
(Tyson et al., 2002). These processes, while being complicated,
have been well studied for fission yeast {\em S. Pombe} and
successful ODE models (Novak et al., 2001) exist.
We choose the most widespread version of the model
(Novak et al., 2001), as often cited in textbooks, which will
serve us as a starting point for our study.

The paper is organized as follows. In Section 2 we show
the passage from the ODE system of algebraic differential equations
for the fission yeast cell cycle, to the limit of the corresponding Boolean
model which we construct. Here also the
difficulties that one can meet when works with Boolean approaches are
discussed. Section 3 explores the dynamics of the derived
Boolean model of the fission yeast cell-cycle.

Finally, in the discussion section the properties of the obtained system
are recapitulated and the Boolean and ODE approaches are compared.

\section{Boolean variables as stationary states}
The passage from a differential equations model to a Boolean model
requires the mapping of continuous solutions (Novak et al, 2001)
into the ON/OFF states of a Boolean network's nodes.
In order to achieve this, the time evolution of a function,
determined by the rate functions and kinetic constants, has to be
replaced with a discrete mapping of the node set into itself.
Moreover, the rules of this self-mapping have to be governed
by logical functions, connecting the binary states of interacting
nodes. The dynamics of the resulting Boolean network is an
ordered sequence of states of the network nodes, instead of
the continuous time output of the ODE model. Having this in mind,
let us find the conditions, which allow to perform the transition
from a differential equations model to a Boolean system.
In the following, we will first describe the passage of the
continuous variables to discrete states and, in a second step,
construct the logical functions representing the dynamics.

\subsection{Stationary states of ODE system}
The model of the fission yeast cell-cycle (Novak, 2001) is based on
the antagonist interaction of $CDK$-cyclin complex with $APC$ (via
proteins $Ste9$ and $Slp1$) and $CDK$ inhibitor $Rum1$. The CDK-cyclin
complex (Cdc2/Cdc13) is represented by two variables -- $preMPF$ and
$MPF$ (``Maturation Promoting Factor''). Furthermore, helper
molecules (Start Kinase  $SK$, transcription factor $TF$, kinase
$k_{wee}$, phosphatase $k_{25}$, and time-delay enzyme $IE$)
participate in the process. Regulatory interaction between these
macromolecules are described by two types of equations --
differential equations for $Ste9$, $Slp1$, $IEP$, $M$, $SK$, $Rum1$,
$preMPF$ and algebraic equations for $k_{wee}$, $k_{25}$,
$TF$ and $MPF$ variables. Among the first ones, some are
of Michaelis-Menten type ($Ste9$, $Slp1$, $IEP$) and some
are exponential growth ($M$, $SK$) equations. The mass
variable $M$ plays a special role as it parametrizes the time
evolution in the system.
The ODE model (Novak, 2001) uses arbitrary units for
concentrations in all equations, since there are few data of
actual protein concentrations. The kinetic constants deretmine
the right timing of the processes.
Solutions of the system show that the concentrations
of the major proteins in general rise or decrease steeply.

To make the transition to a Boolean system, we first need to rescale
the differential equations such that their solutions assume values
between 0 (inactive) and 1 (maximum activity). This is a first step
towards mapping these variables onto Boolean OFF/ON variables with
values 0 and 1. The rescaling does not change the form of equations,
it only affects the values of kinetic constants. To do this, let us
divide all functions by their respective maximum value. For example,
for $Slp1$ we introduce the new rescaled function $Slp1_1$=$Slp1
Ampl$, where $Ampl=2.1$ is the amplitude of the original solution.
Rescaling all variables except $M$ we obtain:
\begin{small}
\begin{eqnarray}
\frac{d[Cdc13_{T1}]}{dt}&=&k_{1}M_{1}-(k'_{2}+k''_{2}[Ste9_{1}]+k'''_{2}[Slp1_{1}])[Cdc13_{T1}] \\
\frac{d[preMPF_{1}]}{dt}&=&k_{wee_{1}}k_{0}(k''_0[Cdc13_{T1}]-[preMPF_{1}])-k_{25}k'''_{0}[preMPF_{1}]-(k'_2+k''_{2}[Ste9_{1}]+k'''_{2}[Slp1_{1}])[preMPF_{1}]\\
\frac{d[Ste9_1]}{dt}&=&(k'_{3}+k''_{3}[Slp1_{1}])\frac{1-[Ste9_{1}]}{J_{3}+1-[Ste9_1]}-(k'_4[SK_{1}]+k_{4}[MPF_{1}]\frac{[Ste9_{1}]}{J_{4}+[Ste9_{1}]}\\
\frac{[Slp1_{T1}]}{dt}&=&k'_{5}+k''_{5}\frac{[MPF_{1}]^4}{{J_{5}^4}+[MPF_{1}]^4}-k_6[Slp1_{T1}]\\
\frac{[Slp1_1]}{dt}&=&k_7[IEP_1]\frac{[Slp1_{T1}]-[Slp1_1]}{J_7+[Slp1_{T1}]-[Slp1_1]}-k_8\frac{[Slp1_1]}{J_8+[Slp1_1]}-k_6[Slp1_1]\\
\frac{d[IEP_1]}{dt}&=&k_9[MPF_1]\frac{1-k'_9[IEP_1]}{J_9+1-k'_9[IEP_1]}-k_{10}\frac{k'_{9}[IEP_1]}{J_{10}+k'_9[IEP_1]}\\
\frac{d[Rum1_{T1}]}{dt}&=&k_{11}-(k_{12}-k'_{12}[SK_1]+k''_{12}[MPF_1])[Rum1_{T1}]\\
\frac{d[SK_1]}{dt}&=&k_{13}[TF_1]-k_{14}[SK_1]\\
\frac{dM}{dt}&=&\mu{M}\label{M_0}\\[0pt]
[TF_1]&=&G(k_{15}M, k'_{16}+k''_{16}[MPF_1], J_{15},J_{16}) \label{TF_1_0} \\
 k_{wee_{1}}&=&k'_{wee}+(k''_{wee}-k'_{wee})G(V_{awee}, V_{iwee}[MPF_1],J_{awee},
 J_{iwee})\label{k_wee_1_0}\\
k_{25_{1}}&=&k'_{25}+(k''_{25}-k'_{25})G(V_{a25}[MPF_1], V_{i25},
J_{a25}, J_{i25}) \label{k_25_1_0}\\[0pt]
[MPF_1]&=&\frac{(k_{17}[Cdc13_{T1}]-k'_{17}[preMPF_1])([k_{17}[Cdc13_{T1}]-k''_{17}[Trimer])}{k'''_{17}[Cdc13_{T1}]}\\
Trimer&=&\frac{k_{18}[Cdc13_{T1}][Rum1_{T1}]}{\sigma+\sqrt{\sigma^2
-k'_{18}[Cdc13_{T1}][Rum1_{T1}]}}\\
\sigma&=&k'_{19}[Cdc13_{T1}]+k''_{19}[Rum1_{T1}]+K_{diss}
\end{eqnarray}
\end{small}
 where the Goldbeter-Koshland (GK) (Goldbeter and Koshland, 1981; Novak et al., 2001) function has the following general form:
 \begin{small}
\begin{equation}
\label{GK}
G(a,b,c,d)=\frac{2ad}{b-a+bc+ad+\sqrt{(b-a+bc+ad)^{2}-4ad(b-a)}}\label{G}.
\end{equation}
\end{small}
Square brackets denote the concentrations of their elements.
The subscript 1 marks the rescaled variables with maximum 1.
The new values of parameters are shown in Table 1.

Next we map the continuous solution of the ODE model (Novak et al, 2001)
into the discrete states of a Boolean network's nodes.
Since in a Boolean model there is no continuous time, but rather
a sequence of switching events between two stationary states of the nodes,
one needs to reduce (whereever possible) the initial dynamics of  the system to a
sequence of the evolution of stationary states. For this, one can
use the results of a bifurcation analysis (Novak et al., 2001) of the
transitions during the cell cycle. Thus, there are some
variables ($Ste9$, $Slp1$, $IEP$) that are described by
Goldbeter-Koshland (GK) functions (Novak et al., 2001) in the stationary
state:
\begin{small}
\begin{eqnarray}
[Ste9_1]&=&G(k_{3}'+k''_{3}[Slp1_1],k'_{4}[SK_1]+k_{4}[MPF_1],
J_{3},
J_{4}) \label{ste9_1}\\[0pt]
[IEP_1]&=&1/k'_{9}G(k_{9}[MPF_1],k_{10}, J_{9}, J_{10}) \label
{IEP_1}\\[0pt]
 [Slp1_1]&=&[Slp1_{T1}]G(k_{7}[IEP_1],k_{8},
J_{7}/[Slp1_{T1}],J_{8}/[Slp1_{T1}]) \label{Slp1_1}
\end{eqnarray}
\end{small}

The characteristic properties of the GK function imply that
its variable mainly resides in two limiting states: High and
Low. The transition time between them is short (as the variables $J$
are small). Therefore, we approximate them as Boolean (binary) variables.

It is easy to see that $Slp1_{T1}$ determines only the amplitude and
the smoothness of the transition in (\ref{Slp1_1}),
therefore, we neglect $Slp1_{T1}$  as a first step towards a Boolean
model and write
\begin{small}
\begin{equation}
[Slp1_1]=G(k_{7}IEP_1, k_{8}, J, J). \label{Slp1_3_0}
\end{equation}
\end{small}
The states of the remaining variables can be described by
Golbeter-Koshland functions, as well. For $SK$, while the
equation for this variable is exponential,  one can evaluate
the right-hand part through a GK function with the
stationary solution (Novak et al., 2001):
\begin{small}
\begin{equation}
[SK_1]=(k_{13}/k_{14})[TF]=(k_{13}/k_{14})G(k_{15}M_1,k'_{16}+k''_{16}[MPF_1],J_{15},J_{16}).
\label{SK_1}
\end{equation}
\end{small}
Furthermore, there are three algebraic equations for $TF$
(\ref{TF_1_0}), $k_{wee}$ (\ref{k_wee_1_0}), and $k_{25}$
(\ref{k_25_1_0}) that contain Golbeter-Koshland functions. Here,
again, the two limiting states of the GK function will be related to
the binary ON/OFF version of the corresponding variables in the
Boolean limit.

Finally, we will simplify the functional behavior for the $Cdc13_{T}$, $preMPF$, and $Rum1_{T}$, as well.
Again we want to neglect the exact path of their transitions, keeping the limiting stationary states,
eventually enabling us to take the limit of Boolean functions as a simplified description of the dynamics.
Equations with the following requirements will allow us to take this limit:

a) In a small neighborhood of the switching point, the functional behavior can be approximated by an exponential rise.

b) On the larger interval, it has a stationary solution with the steep transition between
two limiting stationary states.

c) This function converges to the Heaviside function in the limit of steep transition.

In the following we will see that the Michaelis-Menten dynamics
fulfils these requirements, resulting in exact conformity of initial and
final states and permitting a well controlled passage to a Boolean function.
Let us start from the Michaelis-Menten equation

\begin{equation}
\frac{dX}{dt}=k_{1}\frac{1-X}{J_{1}+1-X}-k_{2}\frac{X}{J_{2}+X}\label{dX/dt_1},
\end{equation}
and first check the condition a). Expanding (\ref{dX/dt_1}) in the neighborhood of switching points where
$X<<J_{1,2}<<1$ and keeping leading order terms yields
\begin{equation}
\frac{dX}{dt}=k_{1}-\frac{k_{2}}{J_{2}}X.
\label{DX/DT}
\end{equation}
This is a common equation of exponential growth/decrease.
This allow us to take equations of exponential growth  as an
expansion of (\ref{DX/DT}) in the neighborhood of switching ON/OFF
points. For $Cdc13_{T}$, for example, the equation
\begin{small}
\begin{equation}
\frac{d[Cdc13_{T1}]}{dt}=k_{1}M-(k'_{2}+k''_{2}[Ste9_1]+k'''_{2}[Slp1_{1}])[Cdc13_{T1}]
\label{Cdc13}
\end{equation}
\end{small}
is the expansion of the equation
\begin{small}
\begin{equation}
\frac{d[Cdc13_{T1}]}{dt}=k_{1}M\frac{1-[Cdc13_{T1}]}{J_{1}+1-[Cdc13_{T1}]}-(k'_{2}+k''_{2}[Ste9_1]+k'''_{2}[Slp1_{1}])[Cdc13_{T1}]\frac{[Cdc13_{T1}]}{J_{2}+[Cdc13_{T1}]}.
\label{Cdc13_2}
\end{equation}
\end{small}
For illustration, in Fig.\ 1, this function is compared to the initial $Cdc13_{T1}$.

Condition b) is satisfied as well, since the stationary states of (\ref{dX/dt_1})
are described by a Goldbeter-Koshland function. Validity of condition c) is shown in
the next section.

Thus, for the above equations we have a system of GK-functions which
are responsible for the transitions between stationary states:
\begin{small}
\begin{eqnarray}
[Cdc13_{T1}]&=&G(k_{1}M, k''_{2}[Ste9_1]+k'''_{2}[Slp1_1], J,
J)\label{Cdc13T_2_2_0}\\[0pt]
[preMPF_1]&=&G(k_0 k_{wee}[Cdc13_{T1}],[k_{wee_{1}}]+k''_0
[k_{25_{1}}]+k'_{2}+k''_{2}[Ste9_1]+k'''_{2}[Slp1_1]), J, J)\\
\label{preMPF_2_2_0}
 [Rum1_{T1}]&=&G(k_{11},
k_{12}+k'_{12}[SK_1]+k''_{12}[MPF_1], J, J). \label{Rum1_2_2_0}
\end{eqnarray}
\end{small}
 One can see in Fig.\ 1 that the new and initial functions
start to grow and start to decrease at the same times, respectively.
Note that the obtained substituted equations (\ref{Cdc13T_2_2_0}-\ref{Rum1_2_2_0}) play only a helper role and cannot be directly
applied to the initial system of differential equations. The
complete correspondence of the obtained system to the initial one is
achieved by the limit transition shown in section \ref{B}.

Summarizing all information above we obtain the system of equations
(\ref{TF_1_0}-\ref{k_25_1_0}), (\ref{ste9_1}-\ref{IEP_1}),
(\ref{Slp1_3_0}-\ref{SK_1}), (\ref{Cdc13T_2_2_0}-\ref{Rum1_2_2_0}),
describing the stationary states of the corresponding variables. The
next step is to perform a transition from the continuous functions
to discrete functions. In functions with continuous time, the
stepwise change corresponds to the Heaviside function.

\subsection{Passage to Boolean variables}\label{B}
We now show that there is an exact passage from the function of
Goldbeter-Koshland (Goldbeter and Koshland, 1981) to the indicator
function (Heaviside step function). Let us remark that in (\ref{GK})
the parameters $a$ and $b$ are functional
variables, whereas $c$ and $d$ (in (\ref{TF_1_0}-\ref{k_25_1_0}),
(\ref{ste9_1}-\ref{IEP_1}), (\ref{Slp1_3_0}-\ref{SK_1}),
(\ref{Cdc13T_2_2_0}-\ref{Rum1_2_2_0}) they are denoted as variables
$J$) are usually fixed and small in all equations. The range of
values of $c$, $d$ varies from 0.001 to 0.01 and only one time for
$Slp1$ it takes on the value 0.3. Very small parameters $c$ and $d$
mean that the enzyme-substrate complex is tightly bound and hardly
dissociates. Thereby in (Novak et al., 2001) an assumption is made that
the enzyme-substrate complexes involved are very stable
(Aguda, 2006). For this reason, let us consider the behavior of the
corresponding GK-functions in the limiting case $c \to 0$, $d \to 0$
while $a$ and $b$ take finite values.

First note that, both, numerator and denominator depend on $d$.
Moreover, $a$ is a numerator's factor. At the same time, $c$ appears
in a sum with the finite terms in the denominator, only (except at
the point $b=a$). Therefore, we can assume, without loss of
generality, that $c=0$ and consider below \begin{small}
$$
G(a,b,0,d)=\frac{2ad}{b-a+ad+\sqrt{(b-a+ad)^{2}-4ad(b-a)}}.
$$
\end{small}
 Using the Taylor expansion of the square root in the
denominator and neglecting higher powers of $d$, we obtain for all
points $a\neq b$
\begin{small}
\begin{equation}
G(a,b,0,d)=\frac{2ad}{b-a+ad+|{b-a}|-ad \frac{b-a}{|{b-a}|}}.
\label{GKexpanded}
\end{equation}
\end{small}
There are two possible cases:
\begin{itemize}
\item
a) if $b-a>0$ then $|b-a|=b-a$ and (\ref{GKexpanded}) takes a form
$$
G(a,b,0,d)=\frac{2ad}{b-a};
$$
\item
b) if $b-a<0$ then $|b-a|=-(b-a)$ and (\ref{GKexpanded}) is simply
G(a,b,0,d)=1.
\end{itemize}
This implies that there is a passage to the limit from
Goldbeter-Koshland function $G(a,b,c,d)$ to the Heaviside function
\begin{small}
$\theta(a-b)$:
$$
\lim_{c,d \to 0}=\theta{(a-b)}= \left\{
\begin{array}{cc}
0,&a<b,\\
1,&a>b.
\end{array}
\right.
$$
\end{small}
Thus, in this limit, the variable $a$ plays the role of an activator
input and the variable $b$ is an inhibitor input. The output is
active/inactive (its Boolean value is equal to one/zero)
if the total value of activator inputs is larger/smaller than the
total value of inhibitor inputs. Thus, the Goldbeter-Koshland-function
converges to the Heaviside function in the limit of steep transitions.

\subsection{Logical Boolean functions}
Let us now rewrite the system of equations in the limit of the parameters $J\to 0$
($J_3$, $J_5$, $J_7$, $J_8$, $J_10$, $J_{16}$, $J_{awee}$,
$J_{iwee}$, $J_{i25}$, $J_{a25}$ and the corresponding parameters $J$ in equations (20), (26-28)
equations) as
% mathematically sound formulation of the limit!
\begin{small}
\begin{eqnarray}
% maybe better eqnarray format
[Ste9_1]&=&\theta(k_{3}'+k''_{3}[Slp1_1]-k'_{4}[SK_1]-k_{4}[MPF_1])
\label{Ste9_2} \\[0pt]
[IEP_1]&=&\theta(k_{9}[MPF_1]-k_{10})\label {IEP_1_2} \\[0pt]
[SK_1]&=&(k_{13}/k_{14})\theta(k_{15}M_1-k'_{16}-k''_{16}[MPF_1])\label{SK_2}\\[0pt]
[TF_1]&=&\theta(k_{15}M-k'_{16}-k''_{16}[MPF_1])\\
\label{TF_1_1} [k_{wee_{1}}]&=&k'_{wee}+(k''_{wee}-k'_{wee})
\theta(V_{awee}-V_{iwee}[MPF_1] \label{k_wee_1_4}\\[0pt]
[k_{25_{1}}]&=&k'_{25}+(k''_{25}-k'_{25})\theta(V_{a25}[MPF_1]-V_{i25})\label{k_25_1_4}\\[0pt]
[Cdc13T_1]&=&\theta(k_{1}M - k''_{2}[Ste9_1]-k'''_{2}[Slp1_1])
\label{Cdc13T_2_0}\\ [0pt]
[preMPF_1]&=&\theta(k_{wee}[Cdc13_{T1}]-[k_{wee_{1}}]-[k_{25_{1}}]-k'_{2}-k''_{2}[Ste9_1]-k'''_{2}[Slp1_1])
\label{preMPF_2}\\[0pt] [Rum1_{T1}]&=&\theta(k_{11}-
k_{12}-k'_{12}[SK_1]-k''_{12}[MPF_1] \label{Rum1_2_0}\\[0pt]
[Slp1_1]&=&\theta(k_{7}[IEP_1] - k_{8}). \label{Slp1_3}
\end{eqnarray}
\end{small}
 Let us add two simplifications to the equations
(\ref{SK_2}) and (\ref{k_wee_1_4}-\ref{k_25_1_4}). In equation
(\ref{SK_2}), the coefficient $k_{13}/k_{14}=1$ and thus can be
neglected. In equation (\ref{k_wee_1_4}), $k_{wee_1}$ can have two
possible values: 0.115 and 1. The first one can be reduced to 0
since it does not change the behavior of the system and analogous
for $k_{25}$
\begin{small}
\begin{eqnarray}
[k_{wee_{1}}]&=&\theta( V_{awee}, V_{iwee}[MPF_1], J_{awee},
J_{iwee}) \\ \label{k_wee_2}
 [k_{25 _{1}}]&=&\theta([MPF_1]- V_{i25})\\
\label{k_25_2}
[SK_1]&=&\theta(k_{15}M- k'_{16}-k''_{16}[MPF_1]).
\label{SK_2_2}
\end{eqnarray}
\end{small}
As a result we have a system of ten equations, where all variables, except
$MPF$ and $M$ take values 0 or 1. $MPF$ and $M$  cannot be
described in this formalism. $MPF$ is represented by an algebraic
equation which cannot be reduced to the GK-function. Taking a closer look,
the solution of $MPF$ does not reach a simple stationary state, instead
there are three typical states of $MPF$ in the system ($preMPF$, $Rum1$, $Cdc13$) --
OFF, intermediate and high activation.

1) If $Cdc13=0$ then $MPF=0$, independently of the states of $preMPF$ and $Rum1$.

2) If $Cdc13=1$ and $preMPF=1$ then $MPF=0.14$, independently of the state of $Rum1$.
This corresponds to an intermediate level, where $preMPF$ prevents high excitation.

3) If $Cdc13=1$, $preMPF=0$, and $Rum1=1$,  then $MPF=1$, with its value slightly decreasing to $MPF=0.93$ if $Rum1=0$. This corresponds to a high level of activation, when $MPF$ is activated by $Cdc13$ and this activation is not reduced by $preMPF$.

Let us reformulate these rules in the following. Assume there are
two variables -- $MPF$ and $MPF2$. The first one is activated by
$Cdc13$. For activation of the second variable $MPF2$, one assumes
that  $MPF$ has to be present as a low-level and $preMPF$ should be
inactive. Thereby, one needs to rewrite the system of equations
(\ref{Ste9_2}), (\ref{Cdc13T_2_0}-\ref{preMPF_2}) taking into
account which level of excitation of $MPF$ is crucial for each
particular variable.
\begin{small}
\begin{eqnarray}
[MPF_1]&=&\theta(Cdc13_{T1})\label{MPF_1_1}\\[0pt]
[MPF2]&=&\theta(MPF_1-[preMPF_1])\label{MPF_2_2}\\[0pt]
[IEP_1]&=&\theta([k_{9}[MPF_1]-k_{10}) \\ \label {IEP_1_2_2}
[TF_1]&=&\theta(k_{15}M-k'_{16}-k''_{16}[MPF2_1])
\\ \label{TF_1_1}
[Rum1_{T1}]&=&\theta(k_{11}- k_{12}-k'_{12}[SK_1]-k''_{12}[MPF_1])\\
\label{Rum1_2}
[k_{wee_{1}}]&=&\theta( V_{awee}- V_{iwee}[MPF_1])
\label{k_wee_2}\\[0pt]
[k_{25 _{1}}]&=&\theta([MPF_1]- V_{i25})\\ \label{k_25_2}
 [SK_1]&=&\theta([TF])\\ \label{SK_2_2}
 [Ste9_1]&=&\theta(k_{3}'+k''_{3}[Slp1_1]-k'_{4}[SK_1]-k_{4}([MPF_1]))\label{Ste9_2_2}.
\end{eqnarray}
\end{small}
%Cell mass

Second, as in the model based on differential equations the cell
mass $M$ takes a special role in the present model. The solution
(Novak et al., 2001) treats it during a cell growth as an independent
variable, which is described by an exponential growth function
(\ref{M_0}). Thus, the variable $M$ corresponds to a time in this
system, which drives the evolution between stationary states
(Novak et al, 2001). In the system, $M$ directly influences $Cdc13$ and
$TF$. As soon as $M$ reaches a threshold value, it activates $Cdc13$
and induces the sequence of consecutive transitions between
stationary states. For $TF$ it plays a role of constantly positive
input, $TF$ is always active unless $MPF$ has a high activity.

As a criterium for the end of the cycle, Novak et al.
(Novak et al., 2001) determine when the cell divides by monitoring the
values of the other variables. When these chosen variables have
certain values that indicate the end of the cell cycle, the current
value of $M$ is divided by two manually, as at the end of mitosis
the cell divides into two daughter cells of approximately equal
masses. Subsequently, $M$ continues its exponential growth, again.

Following this strategy, one needs to distinguish $M$ between two
principal different values -- $M$ and $2M$ in the Boolean model.
Here $M$ works at the beginning of the cell cycle as a trigger of
switching events, whereas $2M$ play a role of an indicator for the
end of the cell cycle. Correspondingly, $M$ becomes $2M$ at the end
of mitosis, when $Slp1$, $Ste9$ and $IEP$ all have high
concentrations.

Thus, one can add the following Boolean rule:
\begin{small}
\begin{eqnarray}
M&=&\theta(2M-[Ste9_1][Slp1_1][IEP_1]) \label{M_1_1}\\
2M&=&\theta(M[Ste9_1][Slp1_1][IEP_1]-2M)\label{M_2_2}.
\end{eqnarray}
\end{small}

Thus, we have a system of equations (\ref{MPF_1_1}-\ref{M_2_2}),
where each variable can take values 0 or 1, only. It is easy to
simplify this system, reducing the kinetic coefficients to 0 or 1
and adding thresholds. Consider, for example, $Cdc13_T$. In Table 2,
based on equation (\ref{Cdc13T_2_0}), we show all possible cases for
$Cdc13$.
 In a more compact form, where the kinetic constants are reduced to 1,
 these rules become
 \begin{small}
\begin{equation}
[Cdc13T_1]=\theta(M - [Ste9_1]-[Slp1_1])\label{Cdc13T_2}.
\end{equation}
\end{small}

Repeating the same procedure for all variables, we obtain the
system of equations:
\begin{small}
\begin{eqnarray}
[preMPF_1]&=&\theta(k_{wee}+[Cdc13_{T1}]-1-[k_{25_{1}}]-[Ste9_1]-[Slp1_1]))\\
\label{preMPF_2_3}
[Slp1_1]&=&\theta([IEP_1])\\
\label{Slp1_3_33}
 [TF_1]=&=&\theta([M]+[2M]-[MPF2_1]\\ \label{TF_1_3}
 [IEP_1]&=&\theta([[MPF_2])\\ \label {IEP_1_2_3}
[Rum1_{T1}]&=&\theta(0.5-[SK_1]-[MPF_1])\\ \label{Rum1_3}
[k_{wee_{1}}]&=&\theta( 0.5- [MPF_1])\\ \label{k_wee_3} [k_{25
_{1}}]&=&\theta([MPF_1]- 0.5)\\ \label{k_25_3}
 [SK_1]&=&\theta([TF_1]) \\ \label{SK_2_3}
[Ste9_1]&=&\theta([Slp1_1]-[SK_1]-[MPF_1])\label{Ste9_2_3}\\[0pt]
M&=&\theta(2M+3-[Ste9_1]-[Slp1_1]-[IEP_1]) \label{M_1_1_1}\\
2M&=&\theta(M+[Ste9_1]+[Slp1_1]+[IEP_1]-3-2M)\label{M_2_2_1}
\end{eqnarray}
\end{small}

\section{Boolean model}
We now have a system of algebraic equations
(\ref{Cdc13T_2}-\ref{Ste9_2_3}), which describe the switch-like
transitions between stationary states, plus Boolean equations
(\ref{MPF_1_1}-\ref{MPF_2_2}),(\ref{M_1_1_1}-\ref{M_2_2_1}) for $M$
and $MPF$. Note that in this discrete system, no information about
continuous time is present any more, except the sequence of events.
To obtain this discrete dynamical sequence, we iteratively solve the
system. We start from the known initial conditions (Novak et al., 2001):
$2M=1$, $Slp1_1=1$, $IEP_1$=1, $Ste9_1=1$, $k_{wee1}=1$, with all
other variables being $0$. Following the terminology of Boolean
models, each variable is represented by one node. The network of nodes
is shown in Fig.2. Each node $i$ has only two states, $S_{i}(n)=1$
(active) and $S_{i}(n)=0$ (inactive). The index $n$ is the number of
iterations. The iterative solution of the system has the following
general form:
\begin{equation}
 S_i(n+1)=\theta\left(
 \sum_k T_{ik}S_k(t)+Q_i\right),
 \label{iteration}
\end{equation}
with the Heaviside function
\begin{equation}
\theta(x)=  \left\{
\begin{array}{cc}
1, & x>0  \\
0, & x \leq 0.
\end{array}
\right.
\end{equation}
All nodes are updated synchronously, which corresponds to the
iteration of the full dynamical system. The interaction matrix
$T_{ik}$ and the state vector $Q_{i}$ determine the transition rule
between states.

The specific values of the interactions $T_{ik}$ are determined as
follows. All interactions between a pair of nodes are defined by an
interaction strength given as the elements $T_{ik}$ and an
activation threshold $Q_{i}$. Positive (negative) arguments of the
$\theta$-functions have $T_{ik}=1$ ($T_{ik}=-1)$. $T_{ii}=1$ for
$M$, this rule is true only for $M$ node, since it is described by a
growing exponential function.

The resulting matrix $T_{ik}$ and vector $Q_{i}$ have the following form:
\begin{small}
$$
T_{ik}=\left(
\begin{array}{cccccccccccccccc}
&Cdc13T&preMPF&MPF&MPF2&k25&k_{wee}&M&Slp1& Ste9&TF&SK&2M&IEP& Rum1\\
Cdc13T&0&0&0&0&0&0&1&-1&-1&0&0&0&0&0\\
preMPF&1&0&0&0&-1&1&0&-1&-1&0&0&0&0&0\\
MPF&1&0&0&0&0&0&0&0&0&0&0&0&0&0\\
MPF2&0&-1&1&0&0&0&0&0&0&0&0&0&0&0\\
k25&0&0&1&0&0&0&0&0&0&0&0&0&0&0\\
k_{wee}&0&0&-1&0&0&0&0&0&0&0&0&0&0&0\\
M&0&0&0&0&0&0&1&-1&-1&0&0&1&-1&0\\
Slp1&0&0&0&0&0&0&0&0&0&0&0&0&1&0\\
Ste9&0&0&-1&0&0&0&0&1&0&0&-1&0&0&0\\
TF&0&0&0&-1&0&0&1&0&0&0&0&1&0&0\\
SK&0&0&0&0&0&0&0&0&0&1&0&0&0&0\\
2M&0&0&0&0&0&0&0&1&1&0&0&-1&1&0\\
IEP&0&0&0&1&0&0&0&0&0&0&0&0&0&0\\
Rum1&0&0&-1&0&0&0&0&0&0&0&-1&0&0&0\\
\end{array}
\right)
$$
$$
Q_i^T=\left(
\begin{array}{cccccccccccccc}
0&-1&0&0&-0.5&0.5&3&0&0&0&0&-3&0&0.5\\
\end{array}
\right)
$$
\end{small}
\section{Results of Boolean simulation of the fission yeast cell cycle}
First we run the model described in the previous section with the
initial conditions (Novak et al., 2001). The temporal evolution of the
protein states is presented in Table 3. One can see that the
iterative solution of the system (\ref{MPF_1_1}-\ref{MPF_2_2})
(\ref{Cdc13T_2}-\ref{M_2_2_1}) is
the switching between unstable stationary states, which coincide
with the corresponding evolution in the ODE model. The final state
is a stable stationary state of the system. One notices that the
initial and end states are identical except for the activation of
the nodes $M$ and $2M$.  The update of nodes $M$ and $2M$, keeping
all other nodes in the same states, starts the new cycle. This
cycling of the model is similar to the realization of cycling in the
original model of differential equations.

Let us briefly summarize our coarse-graining strategy that we
followed in this article.

In Fig. 3, we show consecutive abstractions of the model for the
$Ste9$ and $IEP$ variables as an example. For this we first plot the
dynamics of the differential equations model, then the evolution
between the stationary states solutions of the ODE model, and
finally the sequence of states obtained from the iterated Boolean
network model.

In the next step we run the model starting from each of the $2^{15}$
possible initial states. We find that from all initial states $67\%$
flows into one big attractor. This attractor is the same stable
stationary state that one obtains starting with biological
conditions described above.

\subsection {Mutations}
Let us also compare the behavior of mutants.  We model two mutations
-- $Wee^-$ and $Wee^-Cdc25\triangle$ described in (Novak et al., 2001).
In Boolean models one cannot distinguish between reduced activity
and no activity. This is why we model $Wee^-$ as $Wee\triangle$ in
both cases. We run the model starting with wild type initial
conditions, but with deleted nodes $k_{wee}$, and in the second case
$k_{wee}$ and $k_{25}$. In both mutations, the number of steps is
reduced to 12, as compared to 14 in the wild type cell cycle
described in the previous section. This suggests that the cell can
divide at a smaller size than the wild type, where both mutations
are viable. Our results are in accordance with the predictions of
the earlier differential equation model (Novak et al., 2001).

\subsection {Comparison with an existing Boolean model for the fission yeast cell cycle}
It is interesting to compare this model with another Boolean model
for the fission yeast cell cycle (Davidich and Bornholdt, 2008) that was
built on known biochemical interactions between proteins, only. Both
models are quite similar and have the same connections between
homologue nodes. Their dynamics matches the wild type signaling
sequence during the cell cycle. The difference is that in the
current model the nodes $Cdc13$, $preMPF$, $MPF1$, $MPF2$ correspond
to only two nodes Cdc2/Cdc13 and Cdc2/Cdc13* in
(Davidich and Bornholdt, 2008). So in the model
(Davidich and Bornholdt, 2008), the complex Cdc2/Cdc13 can have two
levels of activation - medium and high. The intermediate level
corresponds to sole activation of the Cdc2/Cdc13 node, whereas a
high level of activation is represented by activation of both,
Cdc2/Cdc13 and Cdc2/Cdc13* .The last one plays the role of a
dephosphorylated Cdc2/Cdc13, that closely corresponds to $MPF$ in
(Novak et al., 2001). In the current model, the node $MPF$ was separated
into two nodes $MPF1$ and $MPF2$ as well, to distinguish different
levels of $MPF$ activation.

The formulas responsible for evolution of proteins are similar in a
sense that in both a threshold activation rule is used. In the model
(Davidich and Bornholdt, 2008), proteins remain active if the
corresponding node was not switched-off by other incoming inhibiting
signals. This rule means that if the protein was activated it
requires some other signals to change its state. Whereas in the
current model one needs to have always a positive incoming signal in
order to keep the protein in its active form.

\section{Discussion and Conclusion}
In this paper,  our aim is to make a connection between two
successfully used methods - the ODE and the Boolean models for
predicting properties of a real biological process. In particular,
we show a possible limit transition between the ODE and the Boolean
model for the fission yeast cell cycle. For this  purpose the known
ODE model (Novak et al., 2001) of fission yeast has been chosen.

In order to do the transition, first the differential equations are
rewritten in a such way that the solutions of equations reach 1 in
their maximum values. Then the obtained equations are transformed in
a limiting procedure to Boolean functions. Firstly, Michaelis-Menten
equations and equations with switch-like behavior can be directly
reduced to Boolean functions. Secondly, a set of equations  with
sigmoidal transfer functions can be replaced with Michaelis-Menten
equations without changing the sequence of states through which the
system evolves. Thirdly, there are also some cases that cannot be
reduced to the two previous ones. It happens when the variable is
described by a constantly growing function or a function which has
distinctly different levels. In this case we propose in the Boolean
model to substitute those variables with two labeling intermediate
and high activity of it. Finally, all continuous solutions  of
equations are mapped into ON/OFF states of Boolean network and the
transition between states are described by Boolean functions.

This Boolean model reproduces the results of the initial ODE model
(Novak et al., 2001). In particular, starting with initial conditions as
in (Novak et al., 2001), the system evolves through the same sequence of
states. The second evidence of similar behavior of the ODE and the
Boolean model is the robustness to the initial conditions. The
Boolean model has a dominant attractor ($67\%$), attracting most of
the trajectories, starting from different initial conditions. The
dominant attractor coincides with initial biological conditions of
the system. The ability to model mutations in the Boolean approach
additionally confirms a good correspondence between the ODE and the
Boolean model.

We find that the transition  to a Boolean model is possible for
differential equations, which have monotonic sigmoidal functions
with distinct upper and lower asymptotes. In particular, firstly, in
our case Michaelis-Menten equations are reduced to  S-shaped
GK-functions which have the necessary asymptotes (Glass and Kauffman, 1973;
Glass and Hill, 1998). This function works as a switch
between the  cases when parameters are defined  as the upper or
lower asymptote and the target control function corresponds to the
maximal or basal rate of biochemical processes. Secondly, here
substituting some equations that have monotonic sigmoidal functions
on the right-hand-side with Michaelis-Menten functions, we also
find, that the exact form of the sigmoidal function does not
strongly influence the behavior of the system. The comparison of the
current model with a previous Boolean model for fission yeast
reveals  that they both have a similar set of variables (proteins)
and similar Boolean functions responsible for update.

Our results also confirm the idea that some molecular control
networks are so robustly designed that timing is not a critical
factor (Braunewell and Bornholdt, 2006). In our case it is possible to
reproduce the main results of (Novak et al., 2001) without including
time, but reproducing the right sequence of events. It supports the
idea that the Boolean approach could contain sufficient information.
Thereby one needs less information about the system,  the knowledge
about reactions on the level of activation/inhibition is sufficient,
which eliminates the problem of finding the right kinetic constants.
Another advantage is the low computational cost of Boolean networks.
The problems  one meets working with the Boolean approach are that
it is sometimes difficult to reduce the concentration level of some
proteins only to ON/OFF states. Sometimes  there are intermediate
states of concentration which need to be separated from high
concentration. In this case two methods are possible. One is, as we
implemented it here, to divide this variable into two and to perform
as two different nodes in a system. Doing this, one needs to take
into account the differences in influences of this protein when it
has intermediate and high concentration. Another solution for a such
situation could be the introduction of two discrete levels of
concentration that the protein can have, for example 1 for
intermediate and 2 for high concentration, which was used in modeling gene regulatory network model for $Arabidopsis thaliana$ flower development (Espinosa-Soto et al., 2004).

We would like to note that the ODE and the Boolean approach are
both useful methods. The advantage of the ODE approach
is that it provides detailed information about the system at any
given time in contrast to the Boolean method, which reproduces only
the right sequence of events. But the costs for this information are
the following. One needs to have exhaustive information about the
reactions, where the most difficult part is to find the right
kinetic constants. Also it demands more computational costs to find
the solution of the system. One could say that the ODE approach is
appropriate when the system is well studied and it is necessary to
make a detailed study of all reactions that take place. On the other
hand if the task is to understand the main principles of some
process and one has less information, the Boolean approach is very
suitable to use.

\newpage

\begin{figure}
\begin{center}
\includegraphics  [width=13cm]{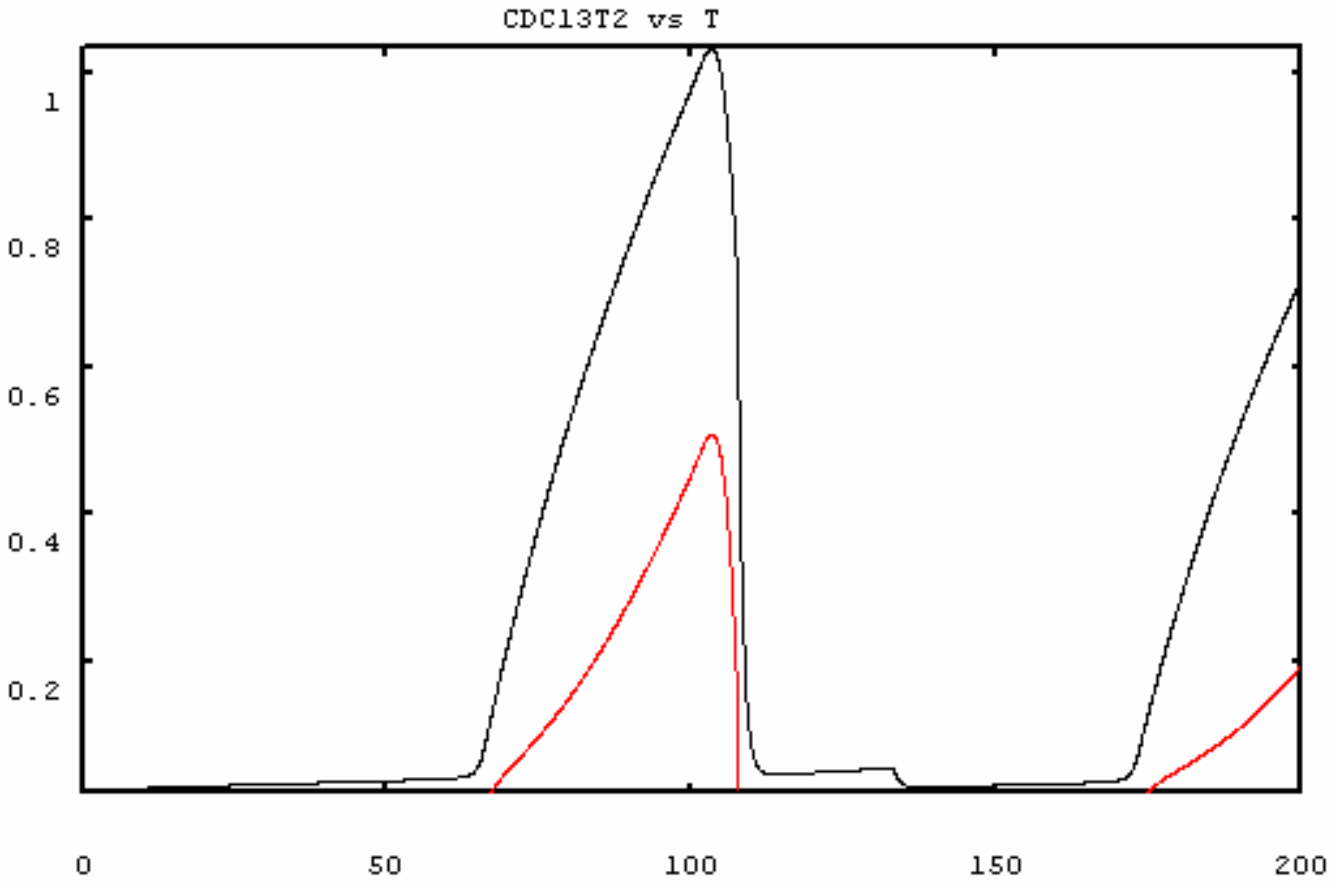}
\vspace{-10cm}\caption {Numerical simulation of (\ref{Cdc13})
(black curve) and (\ref{Cdc13_2}) (red curve). Note that the downward decrease
occurs practically simultaneously.}
\end{center}
\label{cell-cycle}
\vspace{50cm}
\end{figure}

\newpage
\begin{table}
\begin{tabular}{|p{2cm}|p{9cm}|}
\hline  $Cdc13_1$&  $k_{1}=0.04$, $k'_{2}=0.03$, $k''_{2}=1$,
$k'''_{2}=0.21$\\
\hline  $preMPF_1$&  $k'_{0}=1.5$,$k''_{0}=1.17$, $k'''{0}=5$\\
\hline $Ste9_1$& $k'_{3}=1$, $k''_{3}=21$, $J_{3}=0.01$,
$k'_{4}=1.98$, $k_{4}=50.75$\\
\hline $Slp1_{T1}$& $k'_{5}=0.002$, $k'_{5}=0.143$, $k'_{6}=0.048$,
$J_{5}=0.20689$\\
\hline $Slp1_1$& $k_{7}=0.429$, $k_{8}=0.119$, $J_7=0.0005$,
$J_8=0.0005$\\
\hline $IEP$& $k_{9}=0.16$, $J_{9}=0.01$, $k_{10}=0.01$,
$J_{10}=0.011$ $k'_{9}=0.91$\\
\hline $Rum1$ & $k_{11}=0.698$, $k_{12}=0.01$, $k'_{12}=0.99$,
$k''_{12}=4.35$\\
\hline $SK$ & $k_{13}=0.1$, $k_{14}=0.1$\\
\hline $M$ & $\mu=0.005$\\
\hline $TF$ & $k_{15}=3$, $k'_{16}=1$, $k''_{16}=2.9$,
$J_{15}=0.01$, $J_{16}=0.01$\\
\hline $k_{wee}$ & $k'_{wee}=0.115$, $k''_{wee}=1$, $V_{iwee}=1.45$,
$V_{awee}=0.25$, $J_{awee}=0.01$, $J_{iwee}=0.01$\\
\hline $k_{25}$ & $k'_{25}=0.01$, $k''_{25}=1$, $V_{i25}=0.25$,
$V_{awee}=0.36$, $J_{a25}=0.01$, $J_{iwee}=0.01$,
$J_{i25}=0.01$\\
\hline $MPF$ &$k'''_{17}=0.69$, $k_{17}=1.5$, $k'_{17}=1.3$,
$k''_{17}=1.5$,
$k'''{17}=1.5$ \label{MPF1}\\
\hline Trimer & $k_{18}=0.441$, $k'_{18}=0.882$\\
\hline $\sigma$ & $k'_{19}=1.5$, $k''_{19}=0.147$, $K_{diss}=0.001$\\
\hline
\end{tabular}
\caption{Parameter values for the rescaled system of differential equations}
\label{Parameter_ values}
\vspace{50cm}
\end{table}
\vspace{30cm}

\begin{table}
\begin{tabular}{|p{3cm}|p{1.3cm}|p{1.3cm}|p{1.3cm}|p{1.3cm}|}
\hline  Number & $M$ & $Ste9$ & $Slp1$&$Cdc13$\\
\hline 1& 0& 0&0&0\\
\hline 2&1&0&0&1\\
\hline 3& 1&1&0&0\\
\hline 4&1&1&1&0\\
\hline 5&0&1&0&0\\
\hline 6&0&1&1&0\\
\hline 7&1&0&1&0\\
\hline 8&1&1&1&0\\
\hline
\end{tabular}
\caption{Boolean rules for  variable $Cdc13$.}
\vspace{50cm}
\end{table}
\vspace{30cm}

\begin{figure}
\begin{center}
\includegraphics [trim =0mm 100mm 50mm 0mm, scale=0.7]{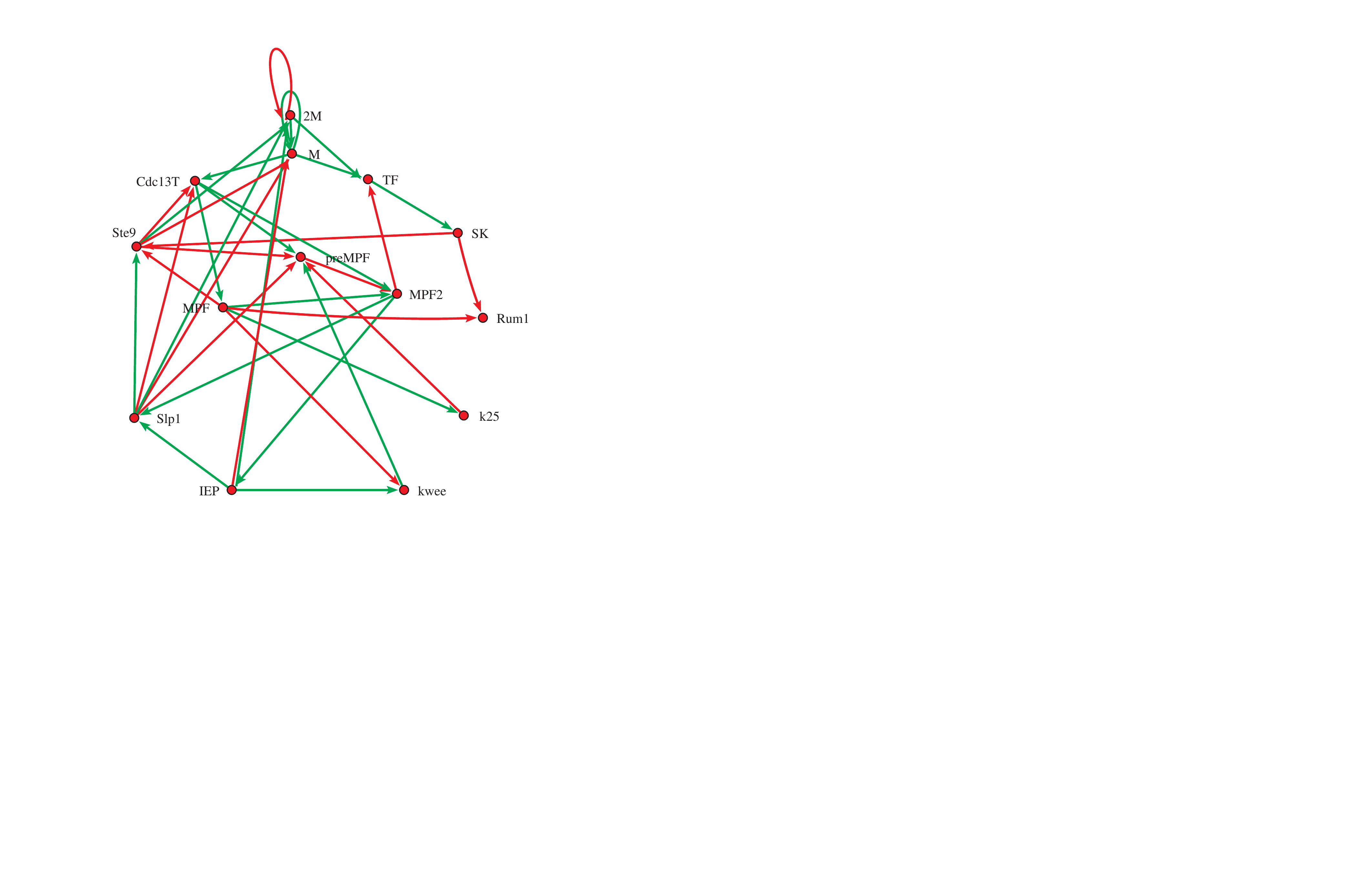}
\vspace{-2cm}
\caption {Interaction network of the Boolean model. Green links correspond to  $T_{ik}$=+1 and red  links to ones to $T_{ik}$=-1}
\end{center}
\label{cell-cycle}
\vspace{50cm}
\end{figure}

\pagebreak
\newpage

\begin{table}
\footnotesize
\begin{tabular}{|p{1.1cm}|p{1.2cm}|p{1.1cm}|p{1.1cm}|p{1.1cm}|p{0.9cm}|p{0.8cm}|p{0.8cm}|p{1cm}|p{1cm}|p{1cm}|p{1cm}|p{1cm}|p{1cm}|p{1cm}|}
\hline  Number of iteration & $Cdc13_{T}$ & $preMPF$ & $MPF1$ & $MPF2$ & $k_{25}$ & $k_{wee}$ & $M$ & $Slp1$ & $Ste9$ & $TF$ & $SK$ &  $2M$ & $IEP$& Rum1\\
\hline  1 &  0 & 0 & 0 & 0 & 0 & 1 & 0 & 1 & 1 &  0 & 0 &
1 & 1 & 1\\
\hline  2  & 0 & 0 & 0 &   0 & 0 &   1 &  1 & 1 &   1 &
1& 0 & 0 &0 & 1 \\
\hline  3 & 0 & 0 & 0 & 0 & 0 & 1 &   1 & 0 &  1 & 0& 1 & 1  & 0 & 1 \\
\hline  4 &  0 & 0 & 0 & 0 & 0 & 1 & 1 & 0 & 0 & 1 & 1 & 0 & 0 & 0 \\
\hline  5 &  1 & 0 & 0 & 0 & 0 & 1 & 1 & 0 & 0 & 1 & 1 & 0 & 0 & 0\\
\hline 6 &   1 & 1 & 1 & 0 & 0 & 1 & 1 & 0 & 0 & 1 & 1 & 0 & 0 & 0\\
\hline   7  & 1 &  1 &   1 & 0  & 1  & 0 & 1 & 0 & 0 & 1&1&0&0&0\\
\hline  8 & 1 & 0 &  1 &  0 & 1 & 0 & 1 & 0 & 0  & 1 & 1 & 0 & 0 & 0\\
\hline  9 &   1 & 0 & 1 & 1 & 1 & 0 & 1 & 0 & 0 & 1 & 0 & 0 & 0
&0 \\
\hline  10 &  1 & 0 & 1 & 1 & 1 & 0 & 1 & 0 & 0 & 0 & 0 & 0 & 1 & 0\\
\hline   11 & 1 & 0 & 1 & 1 & 1 & 0 & 1 & 1 & 0 & 0 & 0& 0
&1&0\\
\hline  12 & 0 & 0 & 1 & 1 & 1 & 0 & 1 & 1 & 0 & 0 & 0 & 0 &
1&0\\
\hline  13 & 0 & 0 & 0 & 1 & 1 & 0 & 1 & 1 & 0 & 0 & 0 & 0 &
1&0\\
\hline  14 & 0 & 0 & 0 & 0 & 0 & 1 & 1 & 1 & 1 & 0 & 0 & 0 &
1&1\\
\hline
\end{tabular}
\caption{Temporal evolution of protein states for the cell-cycle control
network.}
\label{table_temporal}
\vspace{50cm}
\end{table}
\pagebreak
\newpage

\begin{figure}
%\vspace{-13cm}
\begin{center}
\includegraphics [width=10cm] {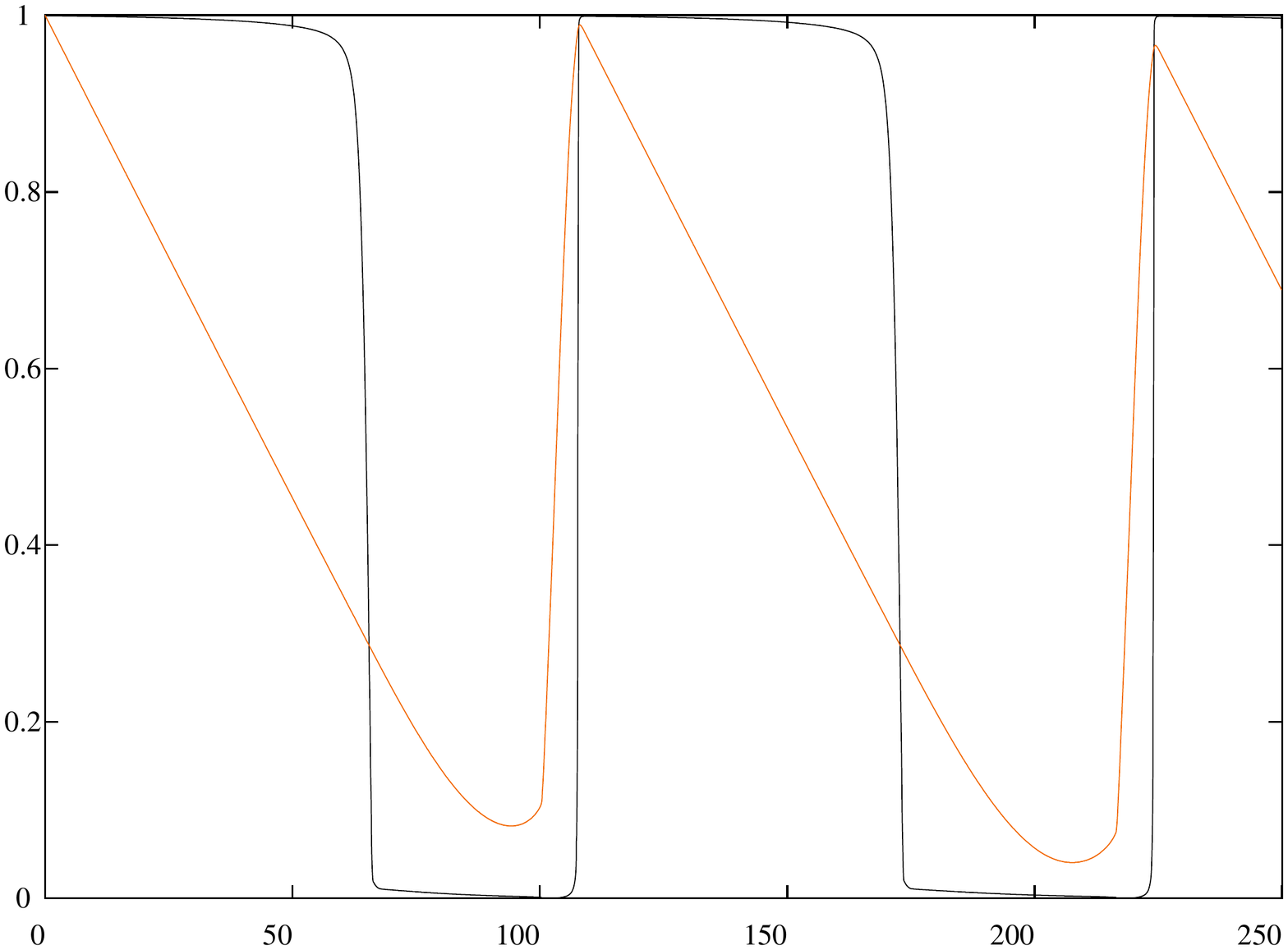}
\includegraphics [width=10cm] {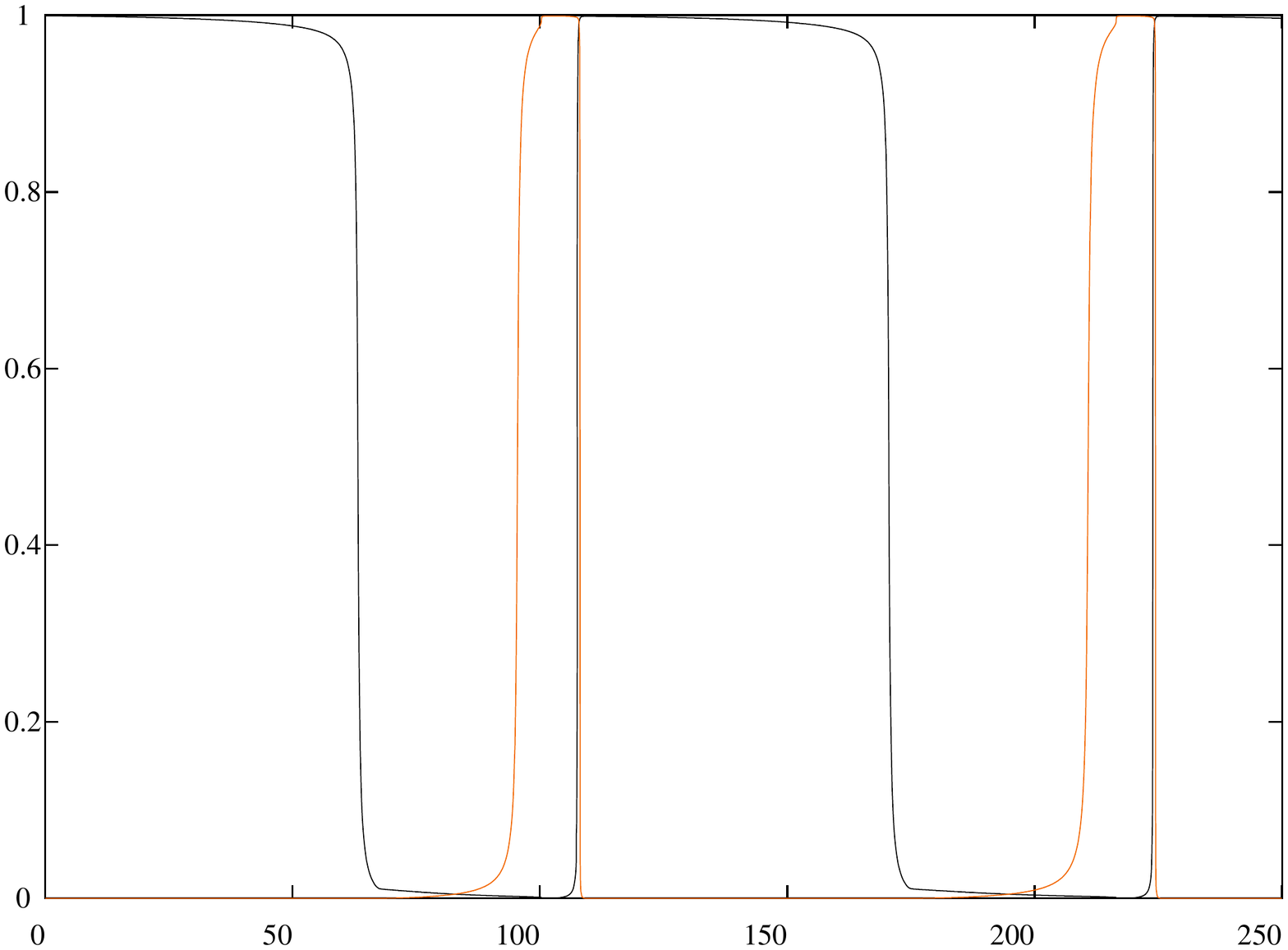}
\includegraphics [width=10cm] {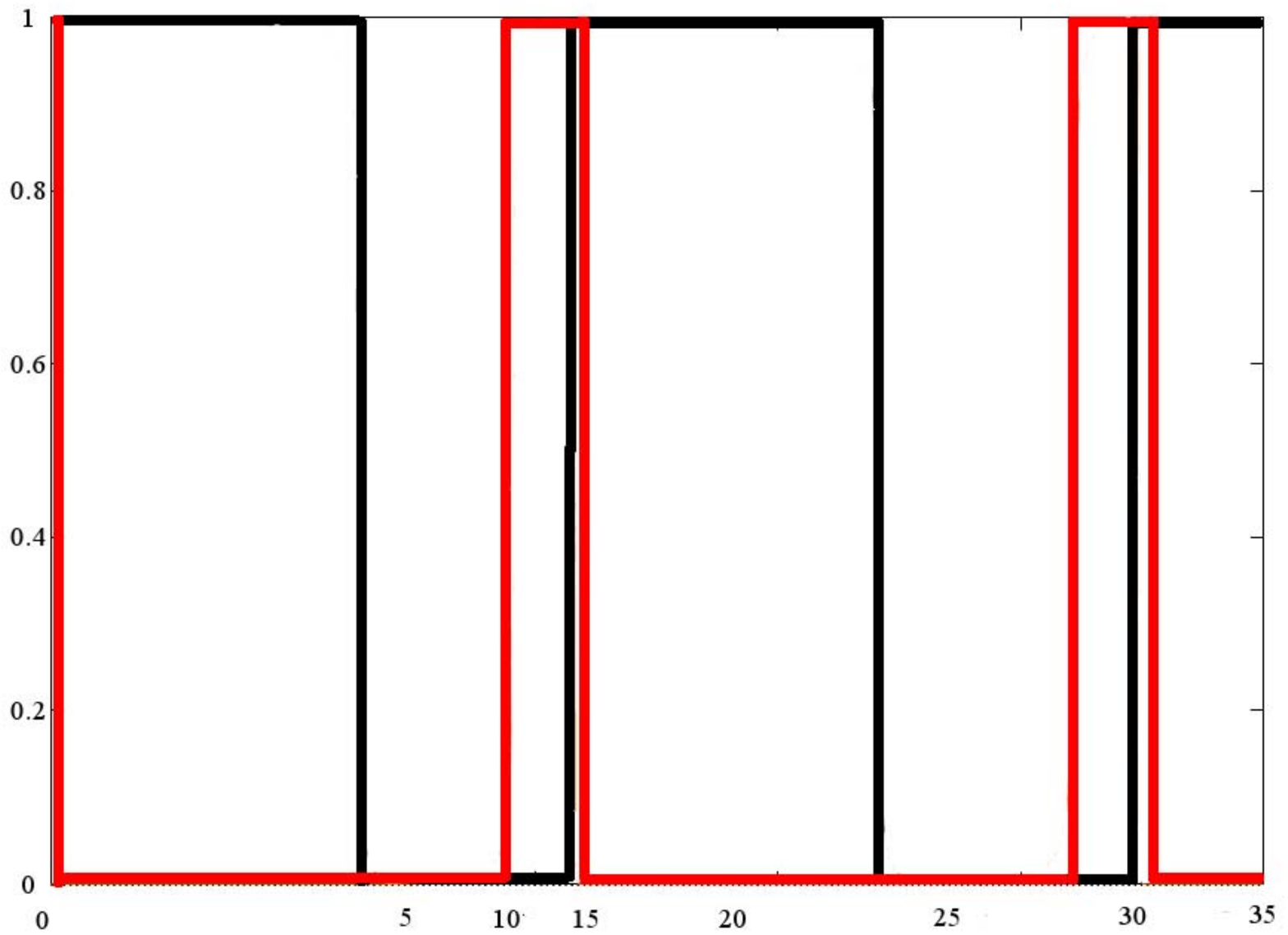}
%\includegraphics [trim =0mm 50mm 0mm 0mm, width=10cm] {IEP_STE_color}
%\includegraphics [trim =0mm 0mm 0mm 0mm, width=10cm] {IEP_STE_state}
%\includegraphics [trim =0mm 30mm 0mm 20mm, width=10cm] {STE_IEP_Bool}

 %\vspace{-19.5cm}
 \caption {a): Dynamics of $IEP$ (red curve) and $Ste9$ (black curve) in the numerical simulation of the system of differential equations. $IEP$ (red curve) and $Ste9$ (black curve).
 b) Numerical simulation of stationary states for $IEP$
 (red curve) and $Ste9$ (black curve).
 c) Boolean networks of dynamical sequence for $IEP$ (red curve) and
 $Ste9$ (black curve).}
%Note that the downward decrease occurs practically simultaneously.}
\end{center}
\end{figure}


\begin{thebibliography}{99}
\bibitem{Aguda2006}
Aguda, D. B., 2006. Modeling the Cell Division Cycle, Lect. Notes Math.
1872, 1-22.

\bibitem{Albert2003}
Albert R., Othmer H. G., 2003. The topology of the regulatory
interactions predicts the expression pattern of the Drosophila
segment polarity genes. J. Theor. Biol. 223, 1-18., doi: 10.1016/S0022-5193(03)00035-3.

\bibitem{Bornholdt2005}
Bornholdt, S., 2005. Systems biology: Less is more in modeling large
genetic networks. Science 310(5747), 449-451.

\bibitem{Braunewell2006}
Braunewell, S., Bornholdt S., 2006. Superstability of the yeast
cell-cycle dynamics: Ensuring causality in the presence of
biochemical stochasticity. J. Theor. Biol. 245(4), 638-643., doi:10.1016/j.jtbi.2006.11.012.

\bibitem{Chen2000}
Chen, K. C., Csikasz-Nagy, A., Gyorffy B., Val, J., Novak, B., Tyson, J. J., 2000. Kinetic
analysis of a molecular model of the budding yeast cell cycle. Mol.
Biol. Cell 11, 369-391.

\bibitem{Davidich_Bornholdt2008}
Davidich, M., Bornholt, S., 2008. Boolean network model predicts cell
cycle sequence of fission yeast. PLOS ONE 27, 3(2): e1672.

\bibitem{Espinosa-Soto2004}
Espinosa-Soto, C., Padilla-Longoria, P., Alvarez-Buylla, E. R., 2004. A
gene regulatory network model for cell-fate determination during
Arabidopsis Thaliana flower development that is robust and recovers
experimental gene expression profiles. Plant Cell 16, 2923-2939.

\bibitem{Faure2006}
Faure, A., Naldi, A., Chaouiya, C., Thieffry, D., 2006. Dynamical analysis
of a generic Boolean model for the control of the mammalian cell
cycle. Bioinformatics 22(14), e124-e131.


\bibitem{Gillespie1976}
Gillespie, D. T., 1976. A general method for numerically simulating the
stochastic time evolution of coupled chemical reactions. J. Comp. Phys.
22, 403-434.

\bibitem{Gillespie1977}
Gillespie, D. T., 1977. Exact stochastic simulation of coupled chemical
reactions. J. Phys. Chem. 81, 2340-2361.

\bibitem{Glass1973}
Glass, L., Kauffman, S. A., 1973. The logical analysis of continuous,
nonlinear biochemical control networks. J. Theor. Biol. 39, 103-129., doi:10.1016/0022-5193(73)90208-7.

%\bibitem{Glass_Kauffman1973}
%Glass, L., Kauffman, SA (1973) The Logical Analysis of Continuous,
%Non-linear Biochemical Control Networks. J. Theor. Biol. 39,
%103-129.

\bibitem{Glass_Hill1998}
Glass, L., Hill, C., 1998. Ordered and
disordered dynamics in random networks. Europhys. Lett., 41(6),
599-604.

\bibitem{GoldbeterKoshland1981}
Goldbeter, A., Koshland, D. E., 1981. An amplified sensitivity arising
from covalent modification in biological systems. Proc. Natl. Acad.
Sci. USA 78, 6840-6844.

\bibitem{Gunsalus2005}
Gunsalus, K. C., Ge, H., Schetter, A. J., Goldberg, D. S., Han J-DJ et al., 2005.
Predictive models of molecular machines involved in Caenorhabditis
elegans early embryogenesis. Nature 436(11), 861-865.

\bibitem{Kaufmann1969}
Kauffman, S. A., 1969. Metabolic stability and epigenesis in randomly
constructed genetic nets. J. Theor. Biol. 22, 437-467., doi:10.1016/0022-5193(69)90015-0.


\bibitem{Li2004}
Li, F., Long, T., Lu, Y., Quyang, Q., Tang, C., 2004. The yeast cell-cycle
network is robustly designed. Proc. Natl. Acad. Sci. U S A 101(14),
4781-4786.

\bibitem{Mendoza1999}
Mendoza, L., Thieffry, D., Alvarez-Buylla, E. R., 1999. Genetic control of
flower morphogenesis in Arabidopsis Thaliana: a logical analysis.
Bioinformatics 15, 593-606.

\bibitem {Novak1993}
Novak, B., Tyson, J.J. 1993. Numerical analysis of a
comprehensive model of M-phase control in Xenopus oocyte extracts
and intact embryos. J. Cell Sci. 106, 1153-1168.

\bibitem{Novak1997}
Novak, B., Tyson, J. J., 1997. Modeling the control of DNA replication in
fission yeast. Cell biology., Proc. Natl. Acad. Sci. U S A 94, 9147-9152.

\bibitem{Novak2001}
Novak, B., Pataki, Z., Ciliberto, A., Tyson, J.J., 2001. Mathematical model
of the cell division cycle of fission yeast. Chaos 11(1), 277-286.


\bibitem{Novak2004}
Novak, B., Tyson, J. J., 2004. A model for restriction point control of
the mammalian cell cycle. J. Theor. Biol. 230,
563-579., doi:10.1016/j.jtbi.2004.04.039.

%\bibitem{Pajek1}
%Nooy W, Mrvar A, Batagelj V (2005) Exploratory Social Network
%Analysis with Pajek, Structural Analysis in the Social Sciences 27.
%Cambridge University Press.

%\bibitem{Pajek2}
%Batagelj V, Mrvar A (1998) Pajek – Program for Large Network
%Analysis. Connections 21(2): 47-57.


\bibitem{Riel2006}
Riel, N. A. W., 2006. Dynamic modelling and analysis of biochemical
networks: mechanism-based models and model-based experiments.
Briefings in Bioinformatics 7(4), 364-374.

\bibitem{Sanchez97}
Sanchez, L., van Helden J., Thieffry, D., 1997. Establishement of the dorso-ventral
pattern during embryonic development of drosophila melanogasater: a logical analysis. J.
Theor. Biol. 21, 189(4), 377-389., doi:10.1006/jtbi.1997.0523.



\bibitem{Sanchez2001}
Sanchez, L., Thieffry, D., 2001. A logical analysis of the drosophila
gap-gene system. J. Theor. Biol. 211, 115-141., doi:10.1006/jtbi.2001.2335.


\bibitem{Smolen2000}
Smolen, P., Baxter, D. A., Byrne, J. H., 2000. Mathematical modeling of gene
networks. Neuron 26, 567-580.

\bibitem{Sveiczer2000}
Sveiczer, A., Csikasz-Nagy, A., Gyorffy, B., Tyson, J. J., Novak, B., 2000.
Modeling the fission yeast cell cycle: Quantized cycle times in
wee1-cdc25 mutant cells. Proc. Natl. Acad. Sci. U S A 97(14), 7865-7870.

\bibitem{Thomas1973}
Thomas, R., 1973. Boolean formalization of genetic control circuits. J. Theor. Biol. 42, 563–585., doi:10.1016/0022-5193(73)90247-6.


\bibitem{Thomas1995}
Thomas, R., Thieffry, D., Kaufmann, M., 1995. Dynamical behaviour of biological regulatory networks. Biological role of feedback loops and practical use of the concept of the loop-characteristic state. Bull Math Biol. 57(2), 247-276.


\bibitem{Thum2003}
Thum, K. E., Shasha D. E., Lejay, L. V., Coruzzi, G. M., 2003. Light- and
carbonsignaling pathways. Modeling circuits of interactions. Plant
Physiol. 132, 440-452.


\bibitem{Tyson2001}
Tyson, J. J., Chen, K. C., Novak, B., 2001. Network dynamics and cell
physiology. Nature Rev. Mol. Cell Biol. 2, 908-916.

\bibitem{Tyson2002}
Tyson, J. J., Csikasz-Nagy, A., Novak, B., 2002. The dynamics of the
cell-cycle regulation. BioEssays 24, 1095-1109.

\bibitem{Tyson2003}
Tyson, J. J., Chen, K. C., Novak, B., 2003. Sniffers, buzzers, toggles and
blinkers: dynamics of regulatory and signaling pathways in the cell.
Curr. Op. Cell Biol. 15, 221-231.





\end{thebibliography}
\end{document}